\documentstyle[aps,prl,multicol,epsf]{revtex} 
\begin{document}
\title{Survival-Time Distribution for Inelastic Collapse}
\author{Michael R. Swift and Alan J. Bray\\} 
\address{
Department of Theoretical Physics,
University of Manchester, Manchester, M13 9PL, UK.}
\maketitle
\begin{abstract}
In a recent publication [PRL {\bf 81}, 1142 (1998)] it was argued
that a randomly forced particle which collides inelastically with
a boundary can undergo inelastic collapse and come to rest 
in a finite time. Here we discuss the survival probability for the
inelastic collapse transition. It is found that the collapse-time 
distribution
behaves asymptotically as a power-law in time, and that the
exponent governing this decay is non-universal. An approximate 
calculation of the collapse-time exponent confirms this behaviour
and shows how inelastic collapse can be viewed as a generalised
persistence phenomenon.
\end{abstract}
\vspace{5mm}

\def\tbar{\overline{t}}

\begin{multicols}{2}
There is currently considerable interest in the statistical
properties of non-equilibrium systems. At the mesoscopic scale,
the dynamics can often be described by Langevin equations\cite{lan},
and much information can be gained from a study of
the first-passage statistics of the underlying stochastic
process\cite{fp}. Quantitative measures such as the survival time or
the persistence probability\cite{pp} have been introduced to characterise
the resistance to temporal fluctuations. In many cases the 
persistence probability is found to decay as a power-law in 
time, and the persistence exponent has been determined, using either exact 
or approximate analytic techniques, for a variety of systems. These include 
the diffusion equation with random initial conditions\cite{dif},
reaction-diffusion systems\cite{rd},
phase-ordering kinetics\cite{pok} and
interfacial growth\cite{ig}. The persistence exponent has also been 
measured experimentally for breath figures \cite{marcos-martin}, a 
2-d liquid crystal system \cite{yurke}, and a 2-d soap froth \cite{tam}.

In a recent publication\cite{nut} it was shown that a randomly
forced particle which collides inelastically with a boundary
can undergo a collapse transition. Namely, if the coefficient of 
restitution is small enough the particle will collide an
infinite number of times and come to rest at the boundary
in a finite time. This transition represents a novel aggregation
mechanism in a driven dissipative system and has applications
ranging from Brownian motion in colloids\cite{col} 
to driven granular media\cite{balls}.
In this note we will address the following
question: given that the particle will collapse at some
finite time $t$, what is the distribution of collapse times $P_c(t)$,
or, alternatively, what is the probability that the particle
will survive up to time $t$. Surprisingly, we find that
the collapse-time distribution has power-law tails and that the
exponent governing the asymptotic decay depends continuously on
the coefficient of restitution. Furthermore, we are able to
make a connection between the collapse transition and the problem of
persistence with partial survival\cite{kill}. Our results show that inelastic
collapse can be viewed as a generalised persistence phenomenon.

We will first outline the arguments which can be used to predict
the collapse transition, as they will form the basis of a systematic
calculation of the collapse-time distribution. Consider a particle
moving in one dimension and subject to a random force. The equation
of motion is
\begin{equation}
\frac{{\rm d}^2 x}{{\rm d}t^2} = \eta(t),\label{eom}
\end{equation}
where $\eta(t)$ is Gaussian white noise with correlator
$\langle \eta(t) \eta(t') \rangle = 2D \delta(t-t')$. 
Whenever it returns to the
origin with speed $v_i$ it is reflected inelastically with a reduced
speed $v_f = r v_i$, $r$ being the coefficient of restitution ($0\le r
\le 1$). The statistics of the motion of the inelastic 
particle can be inferred from
an exact mapping onto an elastic problem, which is statistically equivalent
to a free, randomly accelerated particle. This can be achieved by noting
that the equation of motion, Eq.(\ref{eom}), is invariant under the following
rescaling of variables
\begin{equation}
x\to x'=r^{-3}x;\qquad t\to t'=r^{-2}t,\label{trans1}
\end{equation}
whilst in the primed coordinates the velocity is increased by a factor
$1/r$, $v'=r^{-1}v$. 
Thus, the combination of an inelastic collision followed by a
rescaling of variables results in an elastic collision
in the new coordinates.
If one now performs the rescaling after every
collision, the elapsed time in the inelastic problem can be expressed
as an integral over the effective elastic variables
\begin{equation}
t=\int^{\tbar}_{\tbar_0} \,r^{2n(s)}\, ds,\label{crux}
\end{equation}
where $\tbar$ is the elapsed elastic time and $n(s)$ is the number of
zero crossings up to time $s$ of a free, randomly accelerated particle.
It was shown in \cite{nut} that if $r < r_c = e^{-\pi/\sqrt{3}}$
the integral in Eq.(\ref{crux}) will converge with
probability one as $\tbar \to \infty$.
This is the signature of inelastic collapse, as the particle will collide
an infinite number of times and come to rest at time $t$. 
Here we are interested in the
distribution of $t$ (for $r < r_c$) which is generated by the fluctuations
in $n(t)$.

In Figure 1 we plot the collapse-time distribution
for different values of $r$, as determined from
numerical simulations in which we evolve many trajectories
and form a histogram of the resulting collapse times. The details of the 
simulation technique for the inelastic dynamics
are given in \cite{nut}. It is convenient to
work with the variable $T = \ln t$, and Figure 1 shows that asymptotically,
the probability for collapse within the interval $[T, T+\Delta T]$ decays
exponentially with a rate which depends on $r$. Thus, the collapse-time 
distribution, $P_c(t)$, does indeed show power-law scaling of the form
\begin{equation}
P_c(t) \sim \frac{1}{t^{1+\theta_c(r)}}; \qquad\qquad t \to \infty,
\end{equation}
which serves to define the collapse time exponent $\theta_c(r)$.
Consequently, the survival probability, which is the probability
that the particle has {\em not} collapsed up to time t, will decay 
asymptotically as $t^{-\theta_c (r)}$.

One can argue for the value of $\theta_c$ in two limiting cases. For
$r=0$, the particle will collapse at its first return to the origin
because the reflected velocity will be zero. As the motion up to that
point is just that of a free, randomly accelerated particle, $\theta_c(0)$
can be determined from the first passage distribution of the random 
acceleration process. It is known\cite{fst} that this distribution 
decays asymptotically as $1/t^{5/4}$ which implies $\theta_c(0)=1/4$.
On the other hand, for $r > r_c$ there is no collapse transition, 
so that $\theta_c(r_c)=0$. As can be seen in Figure 1, $\theta_c(r)$ varies
with $r$ between these two limiting values.

\begin{figure}
\narrowtext
\centerline{\epsfxsize\columnwidth\epsfbox{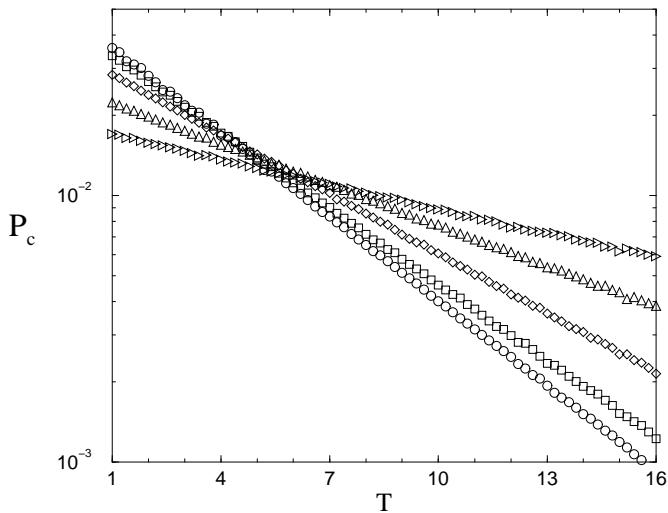}}
\caption{The collapse probability $P_c$ plotted on a semi-log scale
as a function of $T$. Five values of $r$ are shown: $r=0.001, 0.01,
0.04, 0.08, 0.12$ in ascending order on the right of the figure.}
\label{fig1}
\end{figure}

We now turn to an analytic calculation of $\theta_c(r)$ which, although
approximate, does reproduce the simulated values rather accurately.
The starting point is the integral expression for the collapse-time,
Eq.(\ref{crux}), where we have assumed $r < r_c$
and take the limit $\tbar \to \infty$.
The lower limit of the integral and the initial condition of the
particle are arbitrary as we are primarily
interested in the long-time behaviour. 
For calculational clarity it is convenient to set
$\tbar_0=1$ in all that follows\cite{init}.
As $n(t)$ only changes at the zero
crossings of the underlying elastic process, 
the integral, Eq.(\ref{crux}),
can be
expanded as
\begin{equation}
t = t_1 + r^2 (t_2 - t_1) + r^4 (t_3 - t_2) + \,.\,.\,.\,, \label{sum}
\end{equation}
where $t_i$ is the time of the i\,th crossing. For the random acceleration
process, the mean number of crossings up to time $t$ grows logarithmically
with $t$. Therefor, one can map the problem onto a stationary Gaussian
process in the new time variable $T = \ln t$\cite{nut}. 
It is also useful to
express the new crossing times $T_i$ in terms of the intervals between
crossings, $\tau_i$, of the stationary process. One then writes $T_1= \tau_1,
\, T_2= \tau_1+ \tau_2 \,.\,.\,.\,$ and so on. 
Inserting these changes of variable into
Eq.(\ref{sum}), and after rearrangement one finds
\begin{equation}
t = e^{\tau_1}\, [1-r^2 + r^2 e^{\tau_2} \,[1-r^2 + r^2 e^{\tau_3}\,
[ 1 +  \,.\,.\,. \label{hei}
\end{equation}

Next we make an approximation which has been employed in similar 
first-passage and persistence problems\cite{dif,iia}, 
the independent interval approximation
(IIA).
We assume that the $\tau_i$ are all independent and drawn from the
same distribution $P_I(\tau)$, the interval size distribution. Within
this approximation, and exploiting the 
infinite hierarchical structure of Eq.(\ref{hei}),
the stochastic variable $t$ can be expressed as
\begin{equation}
t = e^\tau [ 1 - r^2 + r^2 t'],
\end{equation}
where $t'$ is a random variable drawn from
the same distribution as $t$, $P_c(t)$. It follows that $P_c(t)$ satisfies
the integral equation
\begin{equation}
P_c(t) = \int_1^\infty\!\! dt'\,P_c(t')\int_0^\infty\!\! d\tau\,P_I(\tau)\, 
\delta(t-e^\tau [ 1 - r^2 + r^2 t']). \label{Pc}
\end{equation}
We are interested in the asymptotic behaviour of $P_c(t)$ which we
expect to have the form $P_c(t) \sim 1/t^{1+\theta_c}$. Inserting this
into Eq.(\ref{Pc}) and performing the asymptotic analysis, we are left
with the following equation which $\theta_c$ must satisfy,
\begin{equation}
r^{2\theta_c} \int_0^\infty \, d\tau \, e^{\theta_c \tau} \, P_I(\tau) 
= 1.\label{tc}
\end{equation}
Thus, from knowledge of the interval size distribution
$P_I(\tau)$ one can determine the collapse-time exponent $\theta_c$.

The random acceleration process has been studied by Burkhardt\cite{fst},
and the interval size distribution
has been calculated up to quadrature for arbitrary initial
conditions. Asymptotically 
it is known to decay exponentially, having the form $P_I(\tau)\sim
a_1 e^{-\tau/4}$. The coefficient $a_1$ can be determined exactly
from the results in \cite{fst}. The survival probability, up to time t,
for a particle injected from the origin with speed $u_0$ has the
asymptotic form
\begin{equation}
Q(0, u_0, t) = \frac{3}{\sqrt{2\pi} \Gamma (3/4)} \frac{|u_0|^\frac{1}{2}}
{t^\frac{1}{4}}.\label{q}
\end{equation}
This must be averaged over $u_0$ drawn from the scaling distribution
and correctly weighted to pick out the zeros of $x$; the resulting 
distribution
for $u_0$ is $P(u_0)=2 (u_0/t_0) e^{-u_0^2/t_0}$, 
where $t_0$ is the injection time and $P(u_0)$ is normalised over
the interval $[0,\infty]$. Performing
the average over $u_0$ in Eq.(\ref{q}), one finds that the interval
size distribution has the expected form $a_1 e^{-\tau/4}$, where
$\tau= \ln t/t_0$ and
\begin{equation}
a_1=\frac{3}{4\sqrt{2\pi}} \frac{\Gamma(5/4)}{\Gamma(3/4)} = 0.221314...
\end{equation}
Figure 2 shows $P_I(\tau)$ determined numerically and confirms
the exponential decay at large $\tau$ with the correct value
of $a_1$ given above.

\begin{figure}
\narrowtext
\centerline{\epsfxsize\columnwidth\epsfbox{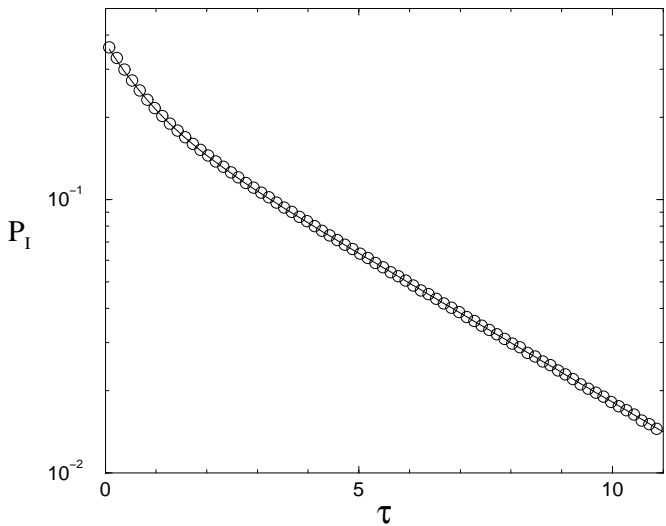}}
\caption{The interval size distribution $P_I(\tau)$ plotted on a semi-log
scale as a function of $\tau$. The points are the numerical data while the
line shows the approximate fit, Eq.(12).}
\label{fig2}
\end{figure}

Knowledge of $P_I(\tau)$ for all $\tau$ is needed to evaluate 
the integral in Eq.(\ref{tc}). A good fit to the numerical data 
can be obtained by writing $P_I(\tau)$ as the sum of just two 
exponentials,
\begin{equation}
P_I(\tau) = a_1 e^{-\theta_1 \tau} + a_2 e^{-\theta_2 \tau},\label{ptau}
\end{equation}
with $\theta_1 = 1/4$ and $a_1=0.2213$.
The parameters $\theta_2$ and $a_2$ can be determined by imposing
two constraints: (1) that $P_I(\tau)$ is normalized over the
interval $[0, \infty]$ and (2) that it predicts the correct density of
zeros\cite{dif} which is governed by the first moment of $P_I(\tau)$, $\rho=
\sqrt{3}/{2 \pi} =1/{\langle\tau \rangle}$. 
These conditions fix the unknown parameters to be $\theta_2 = 1.3254$ and
$a_2 = 0.1521$, and
this approximate expression
is also plotted in Figure 2. It can be seen that 
the parameterised $P_I(\tau)$ fits
the simulation data rather well over the whole range of $\tau$.

When this analytic expression for $P_I(\tau)$ is inserted into
Eq.(\ref{tc})
one finds the following transcendental equation,
\begin{equation}
\frac{1}{r^{2\theta_c}} = \frac{a_1}{\theta_1 - \theta_c} + 
\frac{a_2}{\theta_2 - \theta_c},\label{ans}
\end{equation}
which can be used to determine $\theta_c(r)$.
The numerical solution of this equation is shown in
Figure 3, together with the exponent values measured directly from
the simulations. For $r$ close to $r_c$ it is hard to
access the asymptotic regime numerically, hence the lack of data in this
region of the plot. However,
there is rather good agreement for all the
values of $r$ considered. 
Note that with our parameterised expression for $P_I(\tau)$,
the IIA correctly predicts both the $r\to 0$
limit and the critical value of $r$, namely $r_c=e^{-\pi/\sqrt{3}}$, 
where $\theta_c \to 0$.
Thus, the analytic calculation confirms
that the collapse exponent is non-universal and depends continuously 
on $r$. The inset in Figure 3 shows the behaviour of $\theta_c(r)$ 
for  $r \to 0$, where the approximate analytic calculation and the
simulation data appear to coincide.
To leading order in $r$, Eq.(\ref{ans}) can be expanded to give
\begin{equation}
\theta_c = 1/4 - a_1 \sqrt{r} + .\,.\,.\,.\,.
\end{equation}
As only the first term on the right hand side of Eq.(\ref{ans}) 
contributes in this limit, the expansion is exact within the IIA, 
since both $\theta_1 (=1/4)$ and $a_1$ are known exactly.

\begin{figure}
\narrowtext
\centerline{\epsfxsize\columnwidth\epsfbox{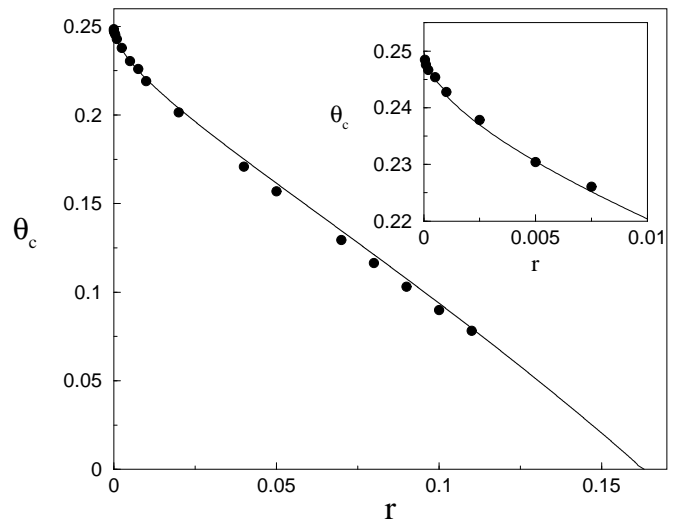}}
\caption{The collapse-time exponent $\theta_c$ as a function of $r$. The
points are the numerical data, obtained from measuring the slopes of
$P_c(T)$. The errors are of the same order as the size of the circles. 
The line is the analytic approximation determined from the
solution of Eq.(13). The inset shows an expanded view of the main
figure for $r$ close to zero where the simulation data and the
approximate calculation appear to coincide.}
\label{fig3}
\end{figure}

Finally we point out an intriguing connection between the exponent 
$\theta_c$ and the persistence exponent of a randomly accelerated  
particle with `partial survival' \cite{kill}. Consider 
the process $x(t)$ described by Eq.(\ref{eom}), and let the 
particle survive with probability $p$ each time its position, $x(t)$, 
crosses through zero. Working in the logarithmic time variable, $T=\ln t$, 
the probability that the particle survives for a time $T$ (in the stationary 
state) is 
\begin{equation}
P_{\rm surv}(T) = \sum_{n=0}^\infty p^n\,P_n(T),
\label{surv}
\end{equation}  
where $P_n(T)$ is the probability that the process has $n$ zero 
crossings in time $T$. We anticipate \cite{kill} that 
$P_{\rm surv}(T) \sim e^{-\theta(p)T}$ for $T \to \infty$, i.e.\ that 
the Laplace transform $\tilde{P}_{\rm surv}(s)$ will have a simple pole at 
$s = -\theta(p)$. 

The Laplace transforms $\tilde{P}_n(s)$ were worked out, in 
terms of the Laplace transform $\tilde{P}_I(s)$ of the interval-size
distribution, $P_I(\tau)$, within the IIA in \cite{dif}. 
Carrying out the sum over $n$ in Eq.(\ref{surv}) 
gives $\tilde{P}_{\rm surv}(s)$, 
and one finds a simple pole given by the condition $p\tilde{P}_I(s) = 1$. 
Setting $s = -\theta(p)$ yields finally 
$p\int_0^\infty d\tau\,e^{\theta(p)\tau}\,P_I(\tau) = 1$ 
as the equation determining $\theta(p)$. This equation has an identical 
form to Eq.(\ref{tc}) for $\theta_c$, suggesting that the collapse 
process with coefficient of restitution $r$ is in some sense equivalent 
to a randomly forced particle with survival probability $p=r^{2\theta_c}$ 
at each collision. As yet, however, we have been unable to establish this 
equivalence outside the IIA.  

To summarize, we have studied the collapse-time distribution, $P_c(t)$, 
for a randomly-forced particle colliding inelastically with a boundary, 
in the regime $r<r_c$ where inelastic collapse occurs.  
We have shown that the distribution has a power-law tail, 
$P_c(t) \sim 1/t^{1+\theta_c}$, characterized by an exponent $\theta_c$ 
which varies continuously with $r$ for $0 \le r \le r_c$. An analytical 
approach, based on the approximation that the intervals 
(in the variable $T=\ln t$) between zero crossings of a free, randomly 
accelerated particle can be treated as statistically independent, 
leads to results which are in good agreement with numerical simulations. 
We have also noted a formal similarity between Eq.(\ref{tc}), which 
determines $\theta_c$ within this approach, and the equivalent equation 
for the persistence exponent of a randomly accelerated particle with 
partial survival \cite{kill}. This similarity invites further study. 
  
This work was supported by EPSRC (UK) under grant number GR/K53208.

\end{multicols}


\begin{references}

\bibitem{lan} P. Langevin, Comptes Rendues {\bf 146}, 530 (1908).

\bibitem{fp} C. W. Gardiner, {\it Handbook of Stochastic Methods
for Physics, Chemistry and the Natural Sciences} (Springer Verlag,
Berlin, 1990).

\bibitem{pp} B. Derrida, A. J. Bray and C. Godrech{\`e}, J. Phys. A
{\bf 27}, L357 (1994).

\bibitem{dif} S. N. Majumdar, C. Sire, A. J. Bray and S. J. Cornell,
Phys. Rev. Lett. {\bf 77}, 2867 (1996); B. Derrida, V. Hakim and
R. Zeitak, ibid., 2871.

\bibitem{rd} J. Cardy, J. Phys. A {\bf 28}, L19 (1995); E. Ben-Naim,
Phys. Rev. E {\bf 53}, 1566 (1996); M. Howard, J. Phys. A {\bf 29},
3437 (1996); S. J. Cornell and A. J. Bray, Phys. Rev. E {\bf 54},
1153 (1996).

\bibitem{pok} B. Derrida, V. Hakim and V. Pasquier, Phys. Rev. Lett.
{\bf 75}, 751 (1995).

\bibitem{ig} J. Krug, H. Kallabis, S. N. Majumdar, S. J. Cornell,
A. J. Bray and C. Sire, Phys. Rev. E {\bf 56}, 2702 (1997).

\bibitem{marcos-martin} M. Marcos-Martin, D. Beysens, J-P. Bouchaud, 
C. Godr\`eche and I. Yekutieli, Physica A {\bf 214}, 396 (1995). 

\bibitem{yurke} B. Yurke, A. N. Pargellis, S. N. Majumdar and C. Sire,
Phys. Rev. E {\bf 56}, R40 (1997).

\bibitem{tam} W. Y. Tam, R. Zeitak, K. Y. Szeto and J. Stavans, Phys. 
Rev. Lett. {\bf 78}, 1588 (1997).

\bibitem{nut} S. J. Cornell, M. R. Swift and A. J. Bray, Phys. Rev. Lett.
{\bf 81}, 1142 (1998).

\bibitem{col} For a review, see T. C. Lubensky, Solid-State-Communications,
{\bf 102}, 187 (1997).

\bibitem{balls} D. R. M. Williams and F. C. MacKintosh, Phys. Rev. E
{\bf 54}, R9 (1996);
M. R. Swift, M. Boamf{\v a}, S. J. Cornell and A. Maritan,
Phys. Rev. Lett. {\bf 80}, 4410 (1998); A. Puglisi, V. Loreto,
U. Marini Bettolo Marconi, A. Petri and A. Vulpiani, Phys. Rev. Lett.
{\bf 81}, 3848 (1998); G. Peng and T. Ohta, cond-mat/9710119.

\bibitem{kill} S. N. Majumdar and A. J. Bray, Phys. Rev. Lett.
{\bf 81}, 2626 (1998).

\bibitem{fst} T. W. Burkhardt, J. Phys. A{\bf 26}, L1157 (1993).

\bibitem{init} We assume that the particle is injected from the
origin at time $\tbar=1$ with a velocity drawn from the scaling
distribution. This assumption will allow us to treat the first
interval on an equal footing with all the rest without changing the
asymptotic behaviour of $P_c(t)$.

\bibitem{iia} J. A. McFadden, IRE Transaction on Information Theory,
IT-4, p.14 (1957).

\end{references}
\end{document}